\documentclass[preprint,preprintnumbers,amsmath,amssymb]{revtex4}

\usepackage{graphicx}
\usepackage{dcolumn}
\usepackage{bm}

\begin{document}


\title{Brane World as a Result of Spontaneous Symmetry Breaking}

\author{Boris E. Meierovich}
\email{meierovich@mail.ru}
\affiliation{P.L.Kapitza Institute for Physical Problems  \\
2 Kosygina str., Moscow 119334, Russia}

 \homepage{http://geocities.com/meierovich}

\date{\today}

\begin{abstract}
The theories of brane world and multidimensional gravity are widely discussed in the literature in connection with problems of evolution of early Universe, including dark matter and energy.
A natural physical concept is that a distinguished surface in the space-time manifold is a topological defect appeared as a result of a phase transition with spontaneous symmetry breaking. The macroscopic theory of phase transitions allows considering the brane world concept self-consistently, even without the knowledge of the nature of physical vacuum.
Gravitational properties of topological defects (cosmic strings, monopoles,...) in extra dimensions are studied in General Relativity considering the order parameter as a vector and  a multiplet in a plane target space of scalar fields. The common results and differences of these two approaches are analyzed and demonstrated in detail.
Among the variety of regular solutions, there are those having brane features, including solutions with multiple branes, as well as the ones of potential interest from the standpoint of the dark matter and hierarchy problems.
Regular configurations have a growing gravitational potential and are able to trap the matter on the brane. If the energy of spontaneous symmetry breaking is high, the attracting potential can have several points of minimum. Identical in the uniform bulk spin-less particles, being trapped within the separate points of minimum, acquire different masses and appear to an observer on brane as different particles with integer spins.

\end{abstract}


\maketitle

\section{\label{sec:level1}Introduction} 

The theories of brane world and multidimensional gravity are widely
discussed in the literature. They continue numerous attempts to find the origin of enormous hierarchy of energy and mass scales observed in nature, to explain the dark matter and dark energy effects, and other long-standing problems in physics. One can find a great variety of separate brane-world models in the literature: thin and thick branes in five and more dimensions, flat or curved branes and bulk, different mechanisms of brane formation, etc.

From my point of view the most natural approach is to consider the appearance of a
distinguished hyper-surface in the space-time manifold as a result of a phase transition with spontaneous
symmetry breaking in the early phase of the Universe formation. The macroscopic Landau theory of phase transitions with spontaneous symmetry breaking allows to consider the brane world concept self-consistently, even without the knowledge of the nature of physical vacuum. It is reasonable to regard the brane world as a topological defect, which inevitably appears as a result of a phase transition with spontaneous symmetry breaking.

The properties of topological defects (strings,  monopoles, ...) are generally described in general relativity with the aid of a multiplet of scalar fields
forming a hedgehog-type configuration in extra dimensions (see \cite{Bron 1} and references there in). The scalar multiplet plays the role of the order parameter. The hedgehog-type multiplet configuration is proportional to a unit vector in the Euclidean target
space of scalar fields. Though this model is self-consistent, it is not the direct way for generalization of a plane monopole to the curved space-time.

In a flat space-time there is no difference between a vector and a hedgehog-type multiplet of scalar fields. On the contrary, in a curved space-time scalar multiplets and real vectors are transformed differently. For this reason in general relativity the two
approaches (a multiplet of scalar fields and a vector order
parameter) give different results which are worth to be compared. It is the subject of this paper. \textit{A priory} it seems more difficult to deal with a vector order parameter, and, probably, it is the reason why I couldn't find in the literature any papers considering phase transitions with a hedgehog-type vector order parameter in general relativity.

The spontaneous symmetry breaking with a hedgehog-type vector order parameter in application to the brane world with two extra dimensions is considered in this report.

\section{General approach}

Let $\phi _{I}$ be a
vector order parameter. Its covariant derivative $\phi _{I;K}$ can be presented as a sum of a symmetric $\phi s_{I;K}$ and antisymmetric $\phi a_{I;K}$ terms:
\begin{equation*}	
\phi _{I;K}=\phi s_{I;K}+\phi a_{I;K},\qquad \phi s_{I;K}=\phi s_{K;I},\qquad \phi a_{I;K}=-\phi a_{K;I}.
\end{equation*}	
The order parameter enters the Lagrangian via scalar bilinear combinations of its covariant derivatives and via a scalar potential $V$ allowing the spontaneous symmetry breaking. A bilinear combination of the covariant derivatives is a tensor
\begin{equation*}
S_{IKLM}=\phi _{I;K}\phi _{L;M}.
\end{equation*}
The most general form of the scalar $S$, formed via contractions of $S_{IKLM},$ is
\begin{equation*}
S=\left( Ag^{IK}g^{LM}+Bg^{IL}g^{KM}+Cg^{IM}g^{KL}\right) S_{IKLM}
\end{equation*}	
where $A,B,$ and $C$ are arbitrary constants. The classification of topological defects with vector order parameters is most visual in terms of the\ symmetric and antisymmetric parts of $\phi _{I;K}.$ In view of
$\phi s_{;K}^{L}\phi a_{L}^{;K}=0,$\quad $\phi a_{;K}^{K}=0$
and
$\phi s_{;K}^{M}\phi s_{;M}^{K}=\phi s_{;K}^{M}\phi s_{M}^{;K},\quad \phi a_{;K}^{M}\phi a_{;M}^{K}=-\phi a_{;K}^{M}\phi a_{M}^{;K}$ the scalar $S$ can be presented in the form
\begin{equation*}
S=A\left( \phi s_{;K}^{K}\right) ^{2}+\left( B+C\right) \phi s_{;K}^{L}\phi s_{L}^{;K}+\left( B-C\right) \phi a_{;K}^{L}\phi a_{L}^{;K}.\qquad \qquad
\end{equation*}
The last term with antisymmetric derivatives is identical to electromagnetism. It becomes clear in usual notations $A_{I}=\phi _{I}/2,\qquad \phi a_{I;K}=\frac{1}{2}\left( \phi _{I;K}-\phi _{K;I}\right) =A_{I;K}-A_{K;I}=F_{IK},$ and $\phi a_{;K}^{L}\phi a_{L}^{;K}=F_{IK}F^{IK}$ is the bilinear combination of the derivatives in electrodynamics. In view of the symmetry of the Christoffel symbols, $\Gamma _{IK}^{L}=\Gamma _{KI}^{L},$ $A_{I;K}-A_{K;I}=\frac{\partial A_{I}}{\partial x^{K}}-\frac{\partial A_{K}}{\partial x^{I}},$ and the combination $F_{IK}F^{IK}$ is free of the derivatives of the metric tensor. On the contrary, the two terms with symmetric covariant derivatives contain not only the components of the metric tensor $g^{IK},$ but also the derivatives $\frac{\partial g_{IK}}{\partial x^{L}}.$
	
The situation with electrodynamics in general relativity is well studied, and we consider below the vector order parameters $\phi _{I}$ with symmetric covariant derivatives: $\phi _{I;K}=\phi _{K;I},$ which is the case of a hedgehog-type topological defect.

The Lagrangian determining gravitational
properties of hedgehog-type topological defects with a vector order parameter is
\begin{equation*}
 L\left(\phi_{I},g^{IK},\frac{\partial g_{IK}}{\partial
x^{L}}\right)=L_{g}+L_{d},\notag
\end{equation*}
where
\begin{equation*}
 L_{g}=\frac{R}{2\kappa^{2}},
 \end{equation*} 
 \begin{equation*}
L_{d}=A\left( \phi _{;K}^{K}\right)
^{2}+B\phi _{;K}^{I}\phi _{I}^{;K}-V\left( \phi ^{K}\phi _{K}\right) .\label{Ldef}
\end{equation*}
$L_{g}$ is the Lagrangian of the gravitational field, $R$ is the
scalar curvature of space-time, $\kappa ^{2}$ is the
(multidimensional) gravitational constant, and $L_{d}$ is the
Lagrangian of a topological defect.

In this report we consider only the case $A=1/2, B=0$, see \cite{JETP} for details. The case $A=0, B=1/2$ is considered in \cite{PRD}.

 Covariant derivation
contain $g^{IK}$
and $\frac{\partial g_{IK}}{\partial x^{L}},$ and for this reason
the Lagrangian has the form
\begin{equation*}
 L\left( \phi _{I},\frac{\partial \phi _{I}}{\partial x^{K}},g^{IK},\frac{\partial g_{IK}}{\partial x^{L}}\right) =\frac{1}{2}\left( g^{IK}\phi _{I;K}\right) ^{2}-V\left( g^{IK}\phi _{I}\phi _{K}\right) ,
 \end{equation*}
 where the symmetry breaking potential $V$ is a scalar function of the scalar $\phi ^{K}\phi _{K}.$ Varying with respect to $\phi _{I},$
 \begin{equation*}
 \frac{1}{\sqrt{-g}}\frac{\partial }{\partial x^{L}}\left( \sqrt{-g}\frac{\partial L}{\frac{\partial \phi _{I}}{\partial x^{L}}}\right) =\frac{\partial L}{\partial \phi _{I}},
 \end{equation*}
 we get the following field equation in covariant form
 \begin{equation*}
\left( \phi _{;A}^{A}\right) ^{;I}=-\frac{\partial V}{\partial \phi _{I}}.
 \end{equation*}
 Substituting $L=L\left( g^{IK},\frac{\partial g_{IK}}{\partial x^{L}}\right) $ into the general formula
 \begin{equation*}
 T_{IK}=\frac{2}{\sqrt{-g}}\left[ \frac{\partial \sqrt{-g}L}{\partial g^{IK}}+g_{MI}g_{NK}\frac{\partial }{\partial x^{L}}\left( \sqrt{-g}\frac{\partial L}{\partial \frac{\partial g_{MN}}{\partial x^{L}}}\right) \right] ,
 \end{equation*}
  we get the following symmetric covariant expression for the energy-momentum tensor
 \begin{equation*}
  T_{IK}=\frac{1}{2}g_{IK}\left( \phi _{;L}^{L}\right) ^{2}+g_{IK}V-2\frac{\partial V}{\partial g^{IK}}-\phi _{I}\left( \phi _{;L}^{L}\right) _{;K}-\phi _{K}\left( \phi _{;L}^{L}\right) _{;I}+g_{IK}\phi ^{L}\left( \phi _{;M}^{M}\right) _{;L}.
\end{equation*}
   Its covariant divergence reduces to
 \begin{equation*}
 T_{I;K}^{K}=-\phi _{I}\left[ \frac{\partial V}{\partial \phi _{K}}+\left( \phi _{;L}^{L}\right) ^{;K}\right] _{;K}.
\end{equation*}
 In view of the field equation we confirm that $T_{I;K}^{K}=0.$ The field equation allows to reduce $T_{IK}$ to a more simple form
 \begin{equation*}
  T_{IK}=\frac{1}{2}g_{IK}\left( \phi _{;L}^{L}\right) ^{2}+g_{IK}V+\left( g_{IL}\phi _{K}-g_{IK}\phi _{L}\right) \frac{\partial V}{\partial \phi _{L}}.
\end{equation*}

We use the Einstein equations
in the form
\begin{equation*} R_{I}^{K}=\kappa
^{2}\widetilde{T}_{I}^{K},
\end{equation*}
where $R_{I}^{K}$ is the Ricci tensor,
and
\begin{eqnarray*} \begin{array}{c}
\widetilde{T}_{I}^{K}=T_{I}^{K}-\frac{1}{d_{0}}\delta _{I}^{K}T
 . \end{array}
\end{eqnarray*}
I omit the detailed derivations (one can find them in my papers \cite{JETP} and \cite{PRD}) and proceed with the results.

\section{Main results}
\begin{table}\caption{Mutual comparison of main features of regular solutions employing a vector order parameter   with the scalar multiplet model. $n_{V}$ is the number of free parameters of the symmetry breaking potential $V(\phi)$.}
  \centering
        $\begin{pmatrix}
      \textbf{Property} & \textbf{Vector}  & \textbf{Scalar multiplet}\\
      $Order  of  Einstein  equations$ & 3 &  4 \\
      $Number of free  parameters$ & n_{V}+1  & n_{V} \\
      $Fine  tuning$ & $\textit{no need}$ & \textit{in some cases} \\

      $Matter trapping$  & yes  & yes \\
      $Presence of $ dV/d\phi $ in equations $& yes  & no \\
     $Presence of $ V $ in  equations$  & no  & yes \\
     $Derivation of $ T_{IK}& wearing  & easy  \\
     $Equations are$ & $\textit{most simple}$  & \textit{simple}\\
     $Strength of gravitational field $ \Gamma& $\textit{arbitrary}$  & \textit{arbitrary}\\
    \end{pmatrix}$
    \label{Table}
\end{table}

\begin{figure}  
\vspace{0cm}
\hspace{-1cm}
   \includegraphics{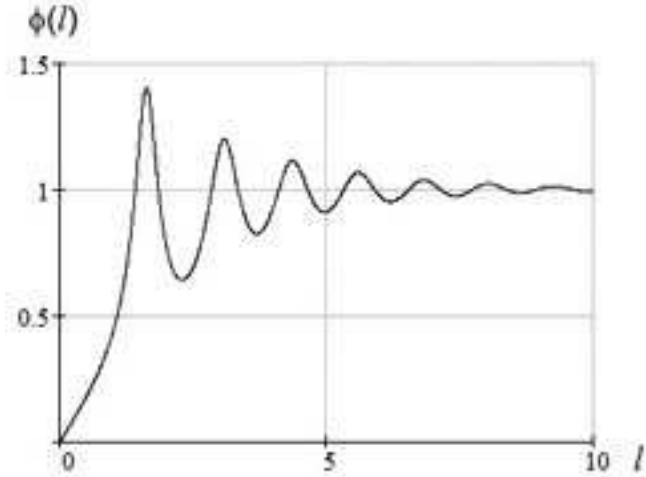}   
   \caption{\label{fig:F1} An example of a solution with oscillating order parameter.}
\end{figure}

The main features of spontaneous symmetry braking with a hedge-hog type vector order parameter in comparison with the widely used previously scalar multiplet model are presented in table \ref{Table}.

\begin{figure}  
\vspace{0cm}
\hspace{-1cm}
   \includegraphics{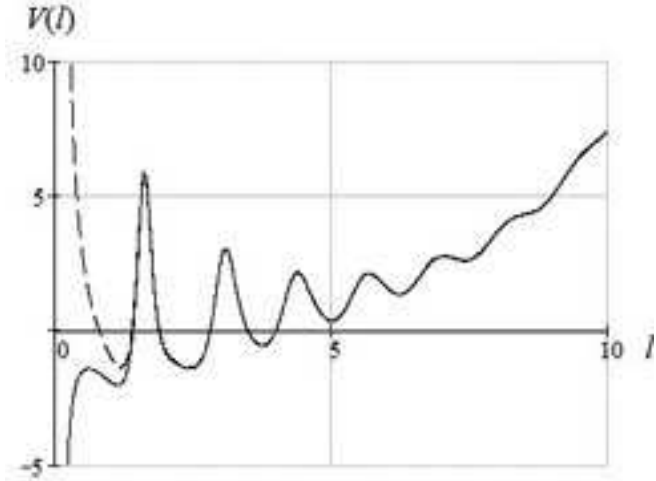}   
   \caption{\label{fig:F2} Gravitational potential 
   for the same set of the parameters as in Fig.\ref{fig:F1}
. The initial mass of a test particle is set $m_{0}=0$. The solid curve corresponds to the extra-dimensional angular momentum $n=0$, and the dashed one -- to $n=1$}
\end{figure}

The solutions have additional parametric freedom. It means that the possibility of existence of the brane world is not connected with any restrictions of fine-tuning type. The origin of the additional parametric freedom is the order of equations, which in case of vector order parameter is less than in scalar multiplet models.

All regular configurations display trapping properties. Among others there are solutions with oscillating order parameter, see an example in Fig.\ref{fig:F1}.  Oscillating behavior of the order parameter gives rise to existence of several points of minimum of the attractive potential (see Fig.\ref{fig:F2}) at different energy levels. Particles,  trapped at different points of minimum,  acquire different masses. If we assume that the length scale   is extremely small, it can be a reason of the observed hierarchy of masses. Integer angular momentum of extra-dimensional motion $n$ is a quantum number. It appears to the observer on brane as an internal momentum of the particle which can be scarcely separated from its spin.

Most elementary particles have half-integer spins. The simple case of spontaneous symmetry breaking, considered above, can not connect the origin of half-integer spins with extra-dimensional angular momenta. Half-integer spins in General Relativity is still a problem.

\end{document}